# High performance magnetic material with Ce and La: an alternative to Nd-Fe-B magnet

Rajiv K. Chouhan,[1,3,*] Arjun K. Pathak,[1] D. Paudyal,[1] and V.K. Pecharsky[1,2]

[1]*The Ames Laboratory, U.S. Department of Energy, Iowa State University, Ames, IA 50011–3020, USA*

[2]*Department of Materials Science and Engineering, Iowa State University, Ames 50011-2300, IA, USA*

[3]*School of Polymers and High Performance Materials, University of Southern Mississippi, Hattiesburg, MS 39406, USA*

A systematic study of magnetocrystalline anisotropy is performed for R(La/Ce/Nd)$_2$Fe$_{14}$B tetragonal compound with the site substitution mechanism. Theoretical calculation suggests the 50% doping with Ce at *4f*-site can lead to competitive magnetic anisotropy to that of the champion magnet Nd$_2$Fe$_{14}$B. Electronic structure calculations are performed using the full-potential linearized augmented plane wave method by inclusion of the spin-orbit coupling and Hubbard (U) interaction in the calculation for the rare-earth elements to get the correct influence of the localized 4f orbitals. Detailed analysis of the magnetic moment and magnetic anisotropy change has been studied by individually inserting the La and Ce atoms at the two inequivalent sites (*4g* and *4f* sites) of the 2-14-B tetragonal structure. Accurate prediction of the total magnetic moment with the orbital contribution in the 2-14-B structure shows the maximum moment for Ce$_2$Fe$_{14}$B (3.86 µ$_B$/f.u. less) compared to Nd$_2$Fe$_{14}$B. Theoretical analysis confirms that regardless of the anti-parallel spin moment emerging in the Ce atom the complex structure of the Ce substituted compound at *4f*-site gives the maximum anisotropy of 2.27 meV/cell with lowering the magnetic moment by 1.26 µ$_B$/f.u. compared to the Nd$_2$Fe$_{14}$B compound.

___________________________
[*]Electronic mail: rajivchouhan@gmail.com; rajiv.chouhan@usm.edu

## I. INTRODUCTION

Permanent magnets can be classified due to their high magnetocrystalline anisotropy energy (MAE), and strong magnetization. Because of their superior properties, permanent magnets have wide range of technological applications such as; they are a key component of wind turbines, electric vehicle motors, smartphones, magnetic resonance imaging (MRI) medical machines and much more. Rare-earth (RE) based permanent magnets have high coercivity which is an important factor to maintain the magnetic stability against the demagnetizing field.[1] Usage of RE ions in the compound enhanced the coercivity through the interaction of their anisotropic localized (4f) electron clouds with the crystal electric field generated by the surrounding charges.[2,3] As a result to the interactions within the compound under the influencing spin-orbit coupling, the magnetic moment of the compound is preferably aligned towards a specific crystalline direction, and therefore, the magnetocrystalline anisotropy energy involved in disrupting the preferred magnetic alignment provides intrinsic stability to the rare earth based magnets. Among the rare-earth based magnets, $Nd_2Fe_{14}B$ is still superior[4] with its unique combination of high magnetic anisotropy and magnetic moment is still the best among other permanent magnets,[5] it is widely known as champion magnet.

Following the discovery of $Nd_2Fe_{14}B$,[4] the understanding of the basic underlying physics about the atomic origin of the magnetic anisotropy within the structure is still incomplete. It is very clear from the previous studies[3,4,5,6,7] that in the 2-14-B structure Nd is the most efficient rare earth element with the dominating axial anisotropy in the $P4_2/mnm$ tetragonal structure, whereas effective magnetic moment comes from the Fe atoms. In-spite of several studies, it is not very clear how different Wyckoff positions (mainly the two non-equivalent sites *4f* and *4g* of rare-earth) in the complex geometry of 2-14-B are responsible in maintaining such a high magnetocrystalline anisotropy. Studies[8,9] also confirmed that the rare earth sites *4f* and *4g* have a potential to be replaced with some abundant rare-earth elements (for example Ce, La, Pr) which can lower the cost of the permanent magnet with less or negligible usage of Nd in the compound. The experimental finding[8] suggest that the Nd site (*4g*) strongly prefers the easy axis [001] direction anisotropy at ambient temperature, however the other *4f*-site has a tendency to reduce the intrinsic stability by aligning partially towards the planar direction. Henceforth, the substitution mechanism can impact the overall criticality to 50% by usage of non-magnetic La atoms or Ce atoms which has recently been reported to perform mix valence states[9,10] of $Ce^{3+}$ and $Ce^{4+}$ with the magnetic *d*-block elements. In addition to this the correct explanation to the total effective magnetic moment (6 $\mu_B$/f.u. of Nd atoms and 31 $\mu_B$/f.u. of Fe atoms) are missing needing to be re-defined. This is mainly due to the spin-orbit coupling contribution associated with the individual atoms in the 2-14-B structure.

In this manuscript, we report the detailed theoretical understanding of the origin of magnetocrystalline anisotropy in the $Nd_2Fe_{14}B$ structure. The importance of sites (Wyckoff positions *4f* and *4g*) is discussed and its overall effect to the magnetic anisotropic performance is calculated by the inclusion of La/Ce with site substitution mechanism. We included the onsite correlation (Hubbard U and J parameter) to capture the better spin-orbit coupling effect in the presence of localized 4f-electrons in the rare-earth atoms. Comprehensive theoretical calculations predict that the Cerium substitution at *4f*-sites in the tetragonal 2-14-B structure is the optimal position to maintain the MAE (reduce by 3.23%) and the total magnetic moment (reduces by 4.5%) which is almost negligible compared to the Nd$_2$Fe$_{14}$B compound. The electron charge density difference analysis confirms the involvement of the geometrical arrangement underlying effect to the hybridization of the B atom in the ab-plane with the *4g*-sites responsible for enhancing the MAE contribution of the Nd atoms along c-axis. This is also confirmed by the experimental work done by D. Haskel*et.al.*[8] for Nd$_2$Fe$_{14}$B compound. The c-axis anisotropy in Nd atoms at *4g*-site is maintained with the overlapped hybridization occuring due to the presence of light weight boron atoms at *4g*-sites which eventually act as a catalyst for localized 4f-electrons of Nd atoms, whereas the insignificant *4f*-site when occupied by Ce atoms a mixed valence state of Ce$^{3+}$ and Ce$^{4+}$ ions is observed and further confirmed by the magnetic moment distribution analysis.

## II. METHODS

We performed the theoretical calculation using the density functional theory (DFT)[11] to study the physical properties of 2-14-B type compounds by the site substitution mechanism at RE(*4f*) and RE(*4g*) Wyckoff positions, respectively. In this study our focus is to see the effect of individual sites in maintaining the high MAE in the tetragonal P4$_2$/*mnm* 2-14-B structure. In addition to the site effects, we performed the doping process on the two non-equivalent sites with the abundant La/Ce atoms in order to reduce the Nd (criticality) usage in the compound. We have analyzed the variation of the MAE, magnetic moments, charge densities, and density of states (DOS) to understand the change of intrinsic magnetic performance. We utilized the full-potential linearized augmented plane wave (FP-LAPW)[12] method, as implemented in the WIEN2k code to study the electronic structure of RE$_2$Fe$_{14}$B. Calculations are performed using generalized gradient approximation (GGA)[13] along with spin-orbit coupling (SOC) and onsite electron correlation (Hubbard U)[14]. The Hubbard parameters used here were U = 6.2 eV and J = 0.7 eV. Around muffin-tin region the spherical harmonics with $l_{max}$=10 were used to expand the wavefunctions, charges and potentials. The plane wave energy cutoff parameters are set as RK$_{max}$ = 7.0 and G$_{max}$ = 12, with the atomic radii for RE (Nd, La, Ce), Fe, and B as 2.5, 2.25 and 1.67a.u., respectively. The k-space integrations have been performed with 6×6×4 Brillouin zone mesh, which was sufficient for the convergence of total energies (10$^{-6}$ Ryd), magnetic moments, and the 4f crystal field (CF) splitting of these 68-atoms per unit cell systems. We used the force method to extract the MAE by taking the difference between the c-axis and the planar total spin polarized eigenvalues energies given by the expression below:

$$MAE \approx \sum_{j=1}^{occupied} \varepsilon_j \vec{b}_{plane} - \sum_{j=1}^{occupied} \varepsilon_j \vec{b}_{z-axis}$$

Where, $\epsilon$ is the sum of eigenvalues for both spins in corresponding directions and $\vec{b}_{c-axis}$ and $\vec{b}_{planar}$ are easy and planar directions. The negative (positive) values for the corresponding MAE relate to the planar and uniaxial anisotropy. The naming convention used for doping is $RE_{(4f)}RE_{(4g)}Fe_{14}B$ throughout the paper.

## III. NUMERICAL RESULTS

$RE_2Fe_{14}B$ type of compounds forms in the complex tetragonal structure with a space group of $P4_2/mnm$,[3] with challenging 68-atoms per unit cell. This structure contains two inequivalent rare earth's (RE) sites namely *4f* and *4g*, surrounded by six inequivalent Fe sites, i.e. Fe(*16k_1*), Fe(*16k_2*), Fe(*8j_1*), Fe(*8j_2*), Fe(*4e*), Fe(*4c*), and one B(*g*) site, respectively as shown in figure-1. The representation of sites can be distinguished by different colors as RE (green as *4f* and blue as *4g*), Fe (pink with numeric values for *16k_1*, *16k_2*, *8j_1*, *8j_2*, *4e*, and *4c*, respectively) and B (black). The lattice parameters used for the calculations are a= 8.80/8.73/8.82 Å and c= 12.20/12.06/12.34 Å for $Nd_2Fe_{14}B$, $Ce_2Fe_{14}B$ and $La_2Fe_{14}B$,[5,15] respectively. We performed the GGA+U+SOC calculations for the $Nd_2Fe_{14}B$ compound to investigate the origin of the c-axis intrinsic magnetocrystalline anisotropy and compared them with the previous available results. Eventually the results reveal that the highest MAE is still governed by the $Nd_2Fe_{14}B$ (2.34 meV/unit cell) among all kinds of doping performed with the Ce/La in the compound. Previous studies[16] have shown that to accurately capture the localized effect of f-orbital in the rare earth elements a proper value of Hubbard U parameter must be included in the theoretical calculations in addition to the spin-orbital contribution. Henceforth, by including all the desired parameters in the calculations the total effective magnetic moment for $Nd_2Fe_{14}B$ is found to be 34.33 $\mu_B$/f.u. which is very close to the earlier results[3] (i.e. 32.5 $\mu_B$/f.u.). Individual spin (orbital) moments for the Nd(*4f*) and Nd(*4g*) sites are 2.81(-1.50) $\mu_B$ and 2.81(-1.50) $\mu_B$, respectively which are similar, whereas significant differences can be seen in the iron moments. The experimental[3] average value of Nd is 2.35$\mu_B$ which is reported by taking the difference between $Nd_2Fe_{14}B$ and $Y_2Fe_{14}B$ per f.u. whereas, one theoretical study[17] reports the average moment of Nd as 2.86(-3.25) $\mu_B$ with the spin orbit coupling. Eventually both the values are inaccurate because taking reference of $Y_2Fe_{14}B$ diminishes the effect of the 4f-orbital contribution in $Nd_2Fe_{14}B$, whereas the theoretical value was done with less k-points (5x5x3), which results in bad converged values. The Fe spin moments for the individual sites are tabulated in Table-1, and compared side by side with the available experimental and theoretical results.[18,19,20] Our calculated spin magnetic moments for Fe atoms are much closer to the experimental values than other previous calculated values. The calculated results are more accurate because of the inclusion of onsite-correlation and spin-orbit coupling effect in the calculations.

To capture the doping effect of La and Ce atoms in 2-14-B compound, we perform the calculation with substitution at *4f* and *4g* sites separately. We found a competitive uniaxial magnetocrystalline anisotropy of 2.26 meV/f.u. for the $CeNdFe_{14}B$ compound compared to 2.34 meV/f.u. for the $Nd_2Fe_{14}B$ compound, when we substitute the *4f* site with the Ce atoms. Surprisingly all substituted structures have shown the uniaxial MAE, which is plotted as a bar graph plot in figure-2. The lowest magnetocrystalline anisotropy is noticed for $La_2Fe_{14}B$ (0.7 meV/f.u.) which is trivial because of the presence of non-magnetic atoms, whereas the La substitution at *4f* and *4g* sites results in a MAE of 1.76 meV/f.u. and 1.63 meV/f.u., respectively. This indicates the importance of the *4g* site over the 4f site in terms of the uniaxial MAE caused by Nd atoms in 2-14-B structure, which is also confirmed by a recent element-specific x-ray magnetic circular dichroism (XMCD) experiment.[8] Close observation of density of states in figure-3 shows that there are two localized peaks below the Fermi energy for the $Nd_2Fe_{14}B$ compound which is mostly coming from the localized 4f-orbital of Nd atoms. A small broadening can be seen in the *4f* site DOS compared to the *4g* site. When we see the individual DOS of Nd atoms for the *4f* and *4g* sites in the presence of a non-magnetic La atom doped case, we notice a sharp peak in Nd atomic DOS. In addition to this, a noticeable low energy shift (left below Fermi energy) in Nd density of state occurs when both *4f* and *4g* sites are occupied by Nd atoms. This is due to the interaction between the Nd atoms of *4f* and *4g* sites in the presence of Fe atoms in its surroundings. In the case of Ce contained 2-14-B compounds, the Ce electron's density of states is always aligned in negative directions to the majority density of states of the compound. This opposite spin alignment was also observed in the other permanent magnetic system $CeCo_5$,[21] which is because of the formation of mixed valence states of $Ce^{3+}$ and $Ce^{4+}$ in the compound. The similar effect of the density of states splitting and broadening in the *4f* site due to presence of Nd occupied *4g* sites can also be seen in CeNdFe$_{14}$B, whereas these phenomena don't occur for Nd atoms at *4g* sites.

We found that the non-magnetic La atom in 2-14-B structure in the doped case compared to the pure (La$_2$Fe$_{14}$B) form has a small magnetic moment (-0.18 and -0.20 µ$_B$) in opposite direction to the total magnetic field direction. Similar trend of magnetic moment alignment of Ce atoms (opposite to total magnetic moment of the cell) are observed in Ce$_2$Fe$_{14}$B, CeNdFe$_{14}$B and NdCeFe$_{14}$B structures. Absolute spin moments of -1.16 and -1.18 µ$_B$ and orbital moments of 0.48 and 0.51 µ$_B$ are found at *4f* and *4g* sites. Hence, the individual total magnetic moment of Ce (*4f* site) and Ce (*4g* site) in Ce$_2$Fe$_{14}$B are the same (i.e. 0.68 µ$_B$) and are opposite to the total magnetic moment 30.47 µ$_B$/f.u. in the unit cell. In spite of having negative moment, the total magnetic moment of the cell is only decreased by a maximum of 1.56 µ$_B$ in Ce substitution at *4f* site among all other doping of La and Ce atoms. This can be understood by considering the Fe moments (table-1) and its density of states (figure-4) trends in the pure compounds i.e. La$_2$Fe$_{14}$B and Ce$_2$Fe$_{14}$B, respectively. Comparing the iron magnetic moment of individual sites of pure La$_2$Fe$_{14}$B and Ce$_2$Fe$_{14}$B structures with respect to the Nd$_2$Fe$_{14}$B structure, we found that there is negligible but small increment in the moment. And because of the presence of 56 Fe-atoms, we cannot see the major effect to the change of total magnetic moments even in the presence of non-magnetic La atoms at RE sites in the 2-14-B system.

We performed the electron difference charge density analysis to reveal the importance of Nd occupied *4g* site in the 2-14-B structure. Figure-5 shows the two-dimensional electron charge difference plots for La$_2$Fe$_{14}$B, Ce$_2$Fe$_{14}$B, Nd$_2$Fe$_{14}$B and CeNdFe$_{14}$B along the ab-plane of the unit cell structure. In all the structures, a strong boron-boron hybridization is observed which is further hybridized with the co-axially aligned Fe(16k$_1$) and Fe(4e) atoms. This unique stable pattern of electrons charge densities results in the uniaxial MAE in all the configurations. The lowest MAE structure (La$_2$Fe$_{14}$B) has no effect on the RE(*4f*) and RE(*4g*) sites because of the presence of the non-magnetic La-atoms, however we can see a noticeable change at the *4g* sites which are situated along the diagonal direction of the boron-boron hybridized pattern. For all cases, Ce$_2$Fe$_{14}$B, Nd$_2$Fe$_{14}$B and CeNdFe$_{14}$B there are strong influences of a diagonal hybridized boron-boron pattern on the *4g* sites, which has introduced the stability in the atomic charge alignment, henceforth making all the structures favorable uniaxial

MAE. Whereas the above influence is missing in *4f*-sites as they lie away from unique diagonal arrangement of boron atoms.

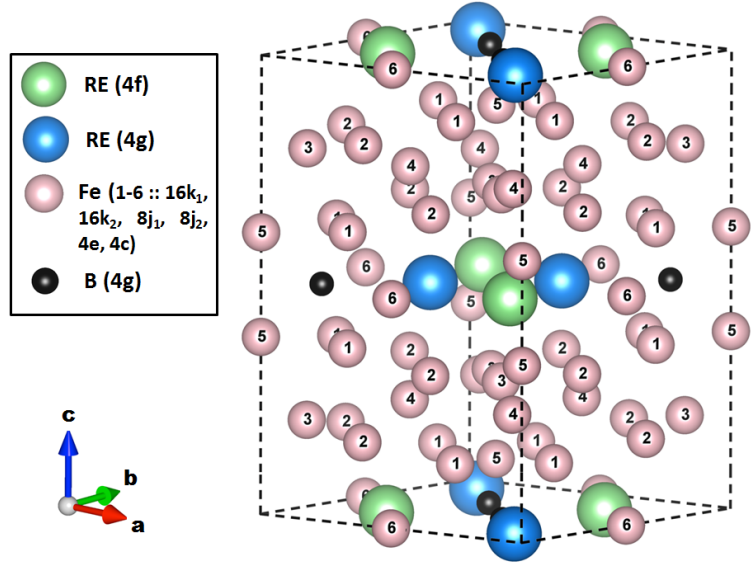

FIG. 1. (Color online) Crystal structures for RE$_2$Fe$_{14}$B having 68 atoms per unit cell and space group P4$_2$/*mnm*.

Table 1. Calculated spin (orbital) magnetic moments for different RE (La/Ce/Nd) and Fe sites, and the total magnetic moment per formula unit (M$_{f.u.}$) in 2-14-B structure (in µ$_B$)

|  | La$_2$Fe$_{14}$B | LaNdFe$_{14}$B | NdLaFe$_{14}$B | Ce$_2$Fe$_{14}$B | CeNdFe$_{14}$B | NdCeFe$_{14}$B | Nd$_2$Fe$_{14}$B | Expt[18] | others[19,20] |
|---|---|---|---|---|---|---|---|---|---|
|  | GGA+U+SOC [ spin (orbital) magnetic moment] in µ$_B$ | | | | | | | | |
| RE(*4f*) | -0.18 (0) | -0.18 (0) | 2.80 (-1.34) | -1.16 (0.48) | -1.16 (0.48) | 2.80 (-1.44) | 2.81 (-1.50) |  | 2.86 (-3.25)[17] |
| RE(*4g*) | -0.20 (0) | 2.80 (-1.49) | -0.19 (0) | -1.18 (0.51) | 2.80 (-1.50) | -1.18 (0.52) | 2.80 (-1.47) |  | 2.85 (-3.08)[17] |
| Fe(*16k$_1$*) | 2.35 | 2.32 | 2.32 | 2.30 | 2.32 | 2.34 | 2.32 | 2.60 | 2.15, 2.45 |
| Fe(*16k$_2$*) | 2.43 | 2.40 | 2.39 | 2.38 | 2.40 | 2.42 | 2.39 | 2.60 | 2.18, 2.43 |
| Fe(*8j$_1$*) | 2.35 | 2.32 | 2.32 | 2.31 | 2.32 | 2.35 | 2.30 | 2.30 | 2.12, 2.33 |
| Fe(*8j$_2$*) | 2.81 | 2.79 | 2.79 | 2.77 | 2.79 | 2.80 | 2.79 | 2.85 | 2.74, 2.72 |
| Fe(*4e*) | 2.22 | 2.20 | 2.20 | 2.16 | 2.20 | 2.21 | 2.21 | 2.10 | 2.13, 2.44 |
| Fe(*4c*) | 2.54 | 2.50 | 2.50 | 2.48 | 2.51 | 2.52 | 2.50 | 2.75 | 2.59, 2.51 |
| B(*4g*) | -0.16 | -0.16 | -0.15 | -0.15 | -0.15 | -0.15 | -0.16 |  |  |
| M$_{f.u.}$ | 32.07 | 33.39 | 33.46 | 30.47 | 32.77 | 33.08 | 34.33 |  |  |

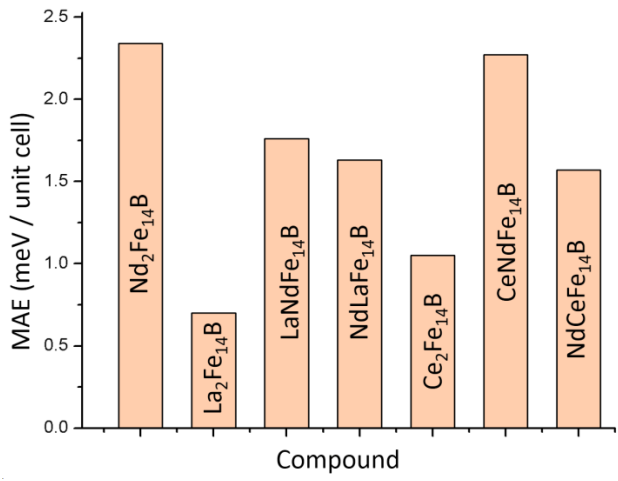

FIG. 2. Magnetocrystalline anisotropy energy (MAE) for various configurations of 2-14-B structures with La/Ce/Nd substitution.

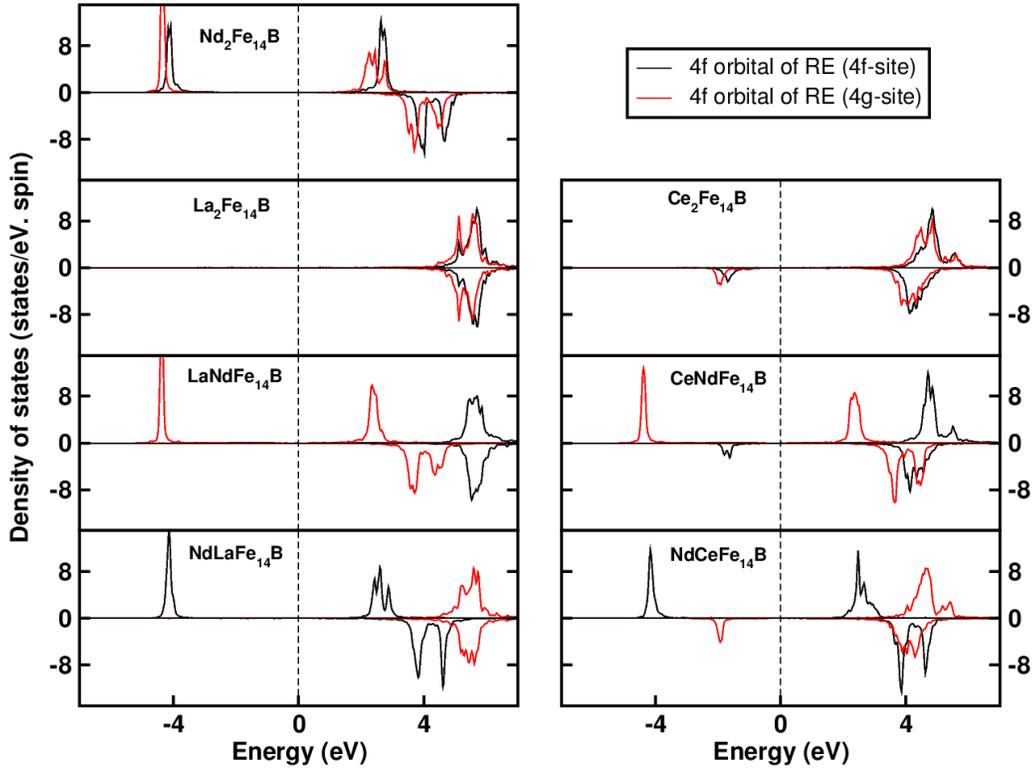

FIG. 3. Calculated site dependent density of states of RE(La/Ce/Nd) atoms for RE$_{(4f)}$ and RE$_{(4g)}$ sites for various configurations in 2-14-B structure.

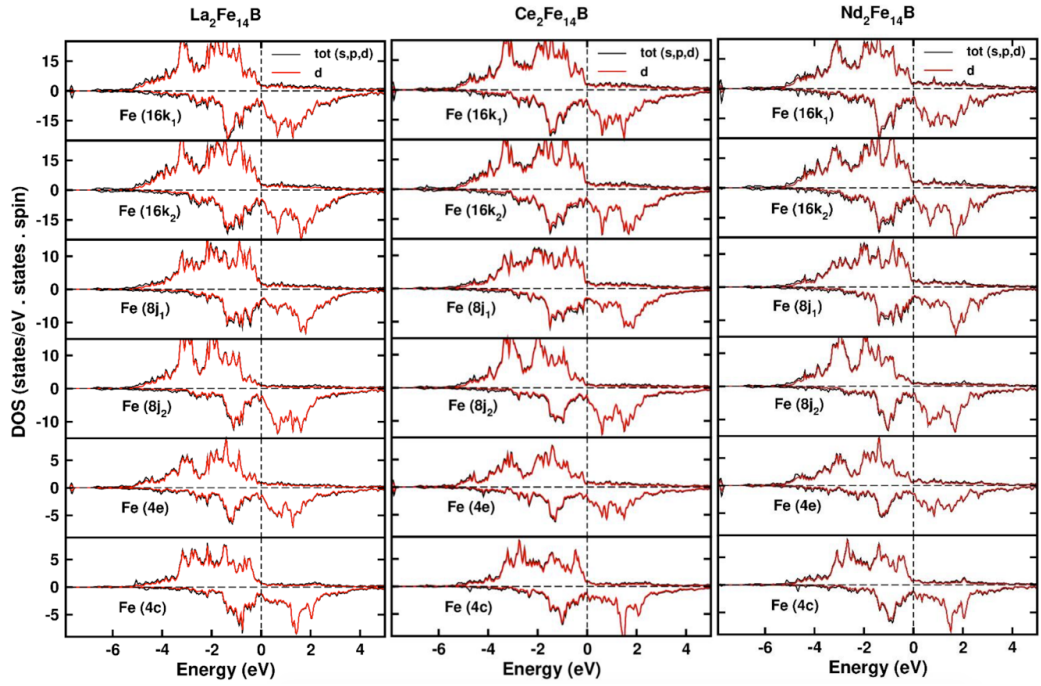

FIG. 4. Calculated site dependent density of states of Fe atoms for $La_2Fe_{14}B$, $Ce_2Fe_{14}B$ and $Nd_2Fe_{14}B$ structures.

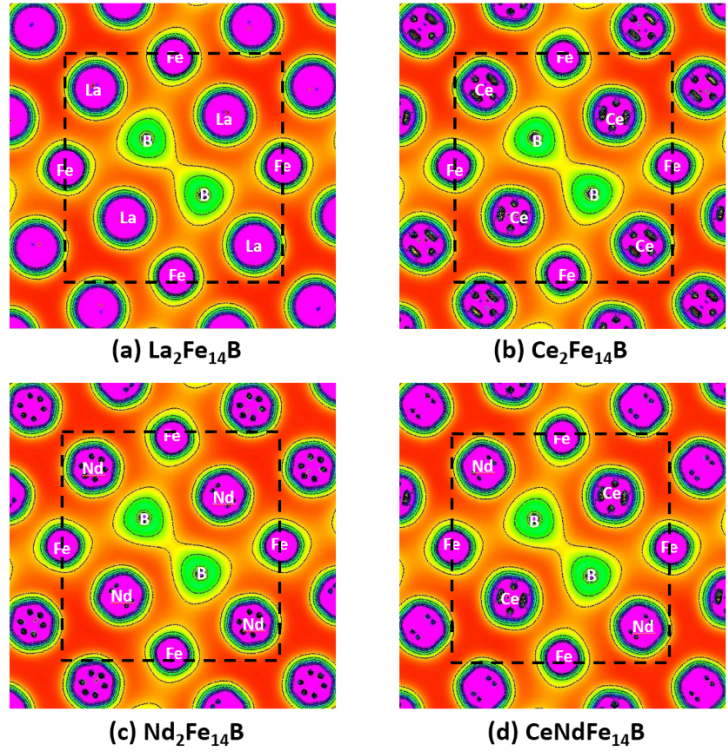

FIG. 5. Two-dimensional electron charge density difference plots for $La_2Fe_{14}B$, $Ce_2Fe_{14}B$, $Nd_2Fe_{14}B$ and $CeNdFe_{14}B$ structure along projected c-axis (ab-plane).

## IV. CONCLUSIONS

In conclusion, we have demonstrated the importance of the RE(*4f*) and RE(*4g*) sites in the 2-14-B structure. We theoretically confirmed that *4g* site is the suitable replacement for the Nd-atoms occupancy to maintain the uniaxial magnetocrystalline anisotropy in the structure. Cerium doping confirms the maximum MAE of 2.26 meV/f.u. when substituted at *4f* sites which is comparable to MAE value of 2.34 meV/f.u. for $Nd_2Fe_{14}B$. The Ce magnetic moment (0.68 $\mu_B$) is oppositely aligned to the Nd atoms, which also confirms the presence of mixed valence state ($Ce^{3+}$ and $Ce^{4+}$) in the 2-14-B structure. The electron charge density difference analysis confirms the involvement of the geometrical arrangement underlying effect to the hybridization of the B atom in the ab-plane with the *4g* sites responsible for enhancing the MAE contribution of the Nd atoms along c-axis.

## V. ACKNOWLEDGEMENT

This work is supported by the Critical Materials Institute, an Energy Innovation Hub funded by the U.S. Department of Energy, Office of Energy Efficiency and Renewable Energy, Advanced Manufacturing Office. The Ames Laboratory is operated for the U. S. Department of Energy by Iowa State University of Science and Technology under contract No. DE-AC02-07CH11358.